\documentclass[10pt,twocolumn,showpacs,longbibliography,amsmath,amssymb,osajnl,floatfix,superscriptaddress,]{revtex4-1}
\usepackage{graphicx}
\usepackage{dcolumn}
\usepackage{bm}
\usepackage{color}
\usepackage{txfonts}
\usepackage{microtype}

\begin{document}

\author{Y.T. Zhao}
\email{zhaoyongtao@mail.xjtu.edu.cn}
\affiliation{MOE Key Laboratory for Nonequilibrium Synthesis and Modulation of Condensed Matter,School of Science, Xi¡¯an Jiaotong University, Xi¡¯an 710049, China}
\affiliation{Institute of Modern Physics, Chinese Academy of Sciences, Lanzhou 730000, China}
\author{Y.N. Zhang}
\affiliation{MOE Key Laboratory for Nonequilibrium Synthesis and Modulation of Condensed Matter,School of Science, Xi¡¯an Jiaotong University, Xi¡¯an 710049, China}
\affiliation{Institute of Applied Physics and Computational Mathematics, Beijing 100088, China}
\author{R. Cheng}
\email{chengrui@impcas.ac.cn}
 \affiliation{Institute of Modern Physics, Chinese Academy of Sciences, Lanzhou 730000, China}
\author{B. He}
\email{hebin-rc@163.com}
 \affiliation{Institute of Applied Physics and Computational Mathematics, Beijing 100088, China}
\author{C.L. Liu}
 \affiliation{Institute of Applied Physics and Computational Mathematics, Beijing 100088, China}
\author{X.M. Zhou}
\affiliation{MOE Key Laboratory for Nonequilibrium Synthesis and Modulation of Condensed Matter,School of Science, Xi¡¯an Jiaotong University, Xi¡¯an 710049, China}
\affiliation{Xianyang Normal University, Xianyang 712000, China}
\author{Y. Lei}
\affiliation{Institute of Modern Physics, Chinese Academy of Sciences, Lanzhou 730000, China}
\author{Y.Y. Wang}
\affiliation{Institute of Modern Physics, Chinese Academy of Sciences, Lanzhou 730000, China}
\author{J.R. Ren}
 \affiliation{MOE Key Laboratory for Nonequilibrium Synthesis and Modulation of Condensed Matter,School of Science, Xi¡¯an Jiaotong University, Xi¡¯an 710049, China}
\author{X. Wang}
 \affiliation{MOE Key Laboratory for Nonequilibrium Synthesis and Modulation of Condensed Matter,School of Science, Xi¡¯an Jiaotong University, Xi¡¯an 710049, China}
\author{Y.H. Chen}
\affiliation{Institute of Modern Physics, Chinese Academy of Sciences, Lanzhou 730000, China}
\author{G.Q. Xiao}
\affiliation{Institute of Modern Physics, Chinese Academy of Sciences, Lanzhou 730000, China}
\author{S.M.Savin}
\affiliation{Alikhanov Institute for Theoretical and Experimental Physics (ITEP) of National Research Center "Kurchatov Institute", Moscow 117218, Russia}
 \author{R.Gavrilin}
\affiliation{Alikhanov Institute for Theoretical and Experimental Physics (ITEP) of National Research Center "Kurchatov Institute", Moscow 117218, Russia}
\author{A.A. Golubev}
\affiliation{Alikhanov Institute for Theoretical and Experimental Physics (ITEP) of National Research Center "Kurchatov Institute", Moscow 117218, Russia}
\affiliation{National Research Nuclear University MEPhI (Moscow Engineering Physics Institute), Moscow 115409, Russia}
\author{D. H.H. Hoffmann}
\affiliation{MOE Key Laboratory for Nonequilibrium Synthesis and Modulation of Condensed Matter,School of Science, Xi¡¯an Jiaotong University, Xi¡¯an 710049, China}
\affiliation{National Research Nuclear University MEPhI (Moscow Engineering Physics Institute), Moscow 115409, Russia}

\title{Significant Contribution of Projectile Excited States to the Stopping of Slow Helium Ions in Hydrogen Plasma}
\date{\today}

\begin{abstract}
The energy deposition and the atomic processes, such as the electron-capture, ionization, excitation and radiative-decays for slow heavy ions in plasma remains an unsolved fundamental problem. Here we investigate, both experimentally and theoretically, the stopping of $100 \,keV/u$ helium ions in a well-defined hydrogen plasma. Our precise measurements show a much higher energy loss than the predictions of the semi-classical approaches with the commonly used effective charge. By solving the Time Dependent Rate Equation (TDRE) with all the main projectile states and for all relevant atomic processes, our calculations are in remarkable agreement with the experimental data. We also demonstrated that, acting as a bridge for electron-capture and ionization, the projectile excited states and their radiative decays can remarkably influence the equilibrium charge states and consequently lead to a substantial increasing of the stopping of ions in plasma.
\end{abstract}

\maketitle



Alpha particle stopping in plasma is a fundamental process in deuteron-tritium fusion scenarios. A precise knowledge of the stopping power is of utmost importance to achieve ignition of the compressed fuel of inertial fusion targets \cite{kodama2001fast,betti2016inertial,hurricane2016inertially,basko2002prospects,moses2009ignition}.
Beam-plasma interactions are also significant for the sciences in astrophysics and applications in plasma stripper in the modern heavy-ion accelerators \cite{mobius1985direct,chabot1995stripping,oguri2000heavy,isenberg2019perpendicular,mccomas2019probing}. The advent of plasma targets at accelerator and high-power-laser facilities initiated a number of experimental activities to study beam plasma interaction phenomena with ionized matter \cite{jacoby1995stopping,belyaev1996measurement,frank2013energy,zylstra2015measurement,sharkov2016high,cayzac2017experimental,braenzel2018charge,frenje2019experimental}.
A reasonably agreement between the theoretical predictions and the experimental data has been reached, especially for fully ionized plasma and heavy ions in the high-energy regime. However, the database for low energy (i.e., with energy of $~10-100 \,keV/u$) heavy-ions propagating in partially ionized plasma is still limited and the physics is not well understood.

Among the problems, universal and yet unsolved, is the influence of excited states of the ion beam on the stopping process. Presently the dynamic evolution of the projectile electronic configuration and its influence on the stopping process is widely ignored or taken into account by an effective charge ($Z_{eff}$), where the projectile is assumed to be in the ground state \cite{northcliffe1960energy,sigmund2001effective,gauthier2013charge,della1987measurement,yan2008study,xu2017determination}. However, excited states are easy to be formed which is almost independent upon the projectile species and plasma states. It is especially prominent in the fusion driven by heavy ions, where most of the electrons by capture and recombination populates in the excited states of the heavy ions \cite{betz1970charge}. They may play important roles in ion stopping since they do affect the projectile charge state and the shielding of the nuclear charge. As it has already pointed out decades before, processes involving excited states may lead to a substantial increase in the stopping cross section for a neutral beam injected into a tokamak plasma \cite{boley1984enhancement}.

Due to the relatively long interaction time and close multi-body collisions in low energy regime, difficulties arise both in theory and in experiment. Even though several models, including ab initio models, have been developed, it is still very difficult to carry out an explicit calculation to describe the whole evolution of the projectile atomic state distribution \cite{betz1972charge,rozet1996etacha,scheidenberger1998charge}. In experiments with discharge plasma and laser-generated plasma, excited states of highly charged projectile ions were taken into account in an indirect way. The effective charge state $Z_{eff}$ was chosen to match the measured energy loss. For instance, Jacoby et.al measured a 35 times higher energy loss in plasma than that in cold gases, and they simply attributed it to a possible effective charge of $6+$ for $Kr$ ions in the plasma \cite{jacoby1995stopping}. More recently, the measurements by Cayzac and his co-workers disproved several standard stopping-power models but supported the theories with a detailed treatment of strong ion-electron collisions,where the effective charge description was used as well \cite{cayzac2017experimental}.Nevertheless, these activities did not reveal details on how the projectile excited states affect their charge state distribution and energy loss.

In this letter, we report the experimental and theoretical investigation of the energy loss for $100 \,keV/u$ helium ion in a hydrogen plasma. The gas discharging plasma is considerably stable and homogenous during the interaction process, which allows a very precise measurement of the stopping power. In order to clarify the contributions of each atomic state, we choose the most simple beam plasma configuration ¨Calpha particles and hydrogen plasma- in which only a handful of atomic-states and charge-transferring channels are involved. More importantly, the alpha particle case is directly relevant to fusion applications.

The experiment was carried out at the $320 \,kV$ highly charged ions platform of HIRFL (Heavy Ion Research Facility in Lanzhou). As shown in Fig. 1, the $100\,keV/u$ $He^{2+}$ ion beam with initial current about $100\,nA$ was collimated into a gas discharging hydrogen plasma, and then recorded by a fast-gated position-sensitive-detector (PSD) behind a $45^{o}$ bending magnet. The beam spot on the PSD was about $1\,mm$, the corresponding energy resolution was $~2\,keV$. The gate width of the PSD was set as $200-500\,ns$, which is triggered with a certain time delay after discharging.
\vspace{-0.5cm}
\begin{figure}[htb] 
\setlength{\abovecaptionskip}{-0.2cm}
\includegraphics[width=8cm]{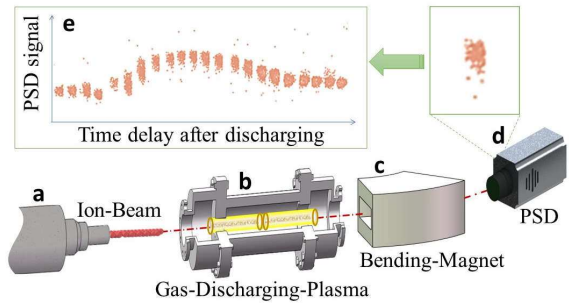}
\caption{\label{fig:wide} Layout of the experiment. The $He^{2+}$ ion-beam from the accelerator ($\mathbf{a}$) was collimated into a gas-discharging-plasma ($\mathbf{b}$) and bended by a magnet ($\mathbf{c}$), then recorded by a fast-gated PSD ($\mathbf{d}$). By sorting the PSD signals in the sequence of the time delay after discharge ($\mathbf{e}$), the temporal profile of energy loss was obtained.}
\setlength{\belowcaptionskip}{-0.2cm}   
\end{figure}

The plasma was generated by igniting an electric discharge in hydrogen gas in two collinear quartz tubes. A high voltage of $2-5\,kV$ is set in between of the two tubes and produces discharging current toward two opposite directions with the grounded electrodes at either end. Such a design can suppress the influence of electromagnetic field caused by the discharging. With an initial gas pressure of 1-5mbar, the discharging can generate plasma with temperature of $1-2\, eV$ and a free electron density of $10^{16}-10^{17} \,cm^{-3}$. The typical lifetime of the plasma is several microseconds, which is two orders of magnitude longer than the ion-plasma interaction time in our measurements (It takes $~ 50\,ns$ for the $100\,keV/u$ helium ion passing though the plasma). The plasma parameters were well characterised by means of Mach-Zehnder interferometer \cite{kuznetsov2013measurements}. A similar plasma device has been successfully used in the previous experiments in higher energy regime \cite{golubev1998dense}. Before this work, we have measured the energy loss of proton with initial energy of $100\,keV$ for different initial gas density and discharging high-voltage, and the measurements showed an excellent agreement with the theoretical predictions both for neutral gas and for plasma target. That means the measured energy loss of the proton can be served as a probe of the target parameters as well \cite{cheng2018energy,chen2018experimental}.

Fig.2 presents the measured energy loss for $100\,keV/u$ helium ions in the plasma at a certain time after discharging at $3\,kV$. At the very beginning of the discharging, the energy-loss in the cold hydrogen gas has been recorded, from the measured energy loss, the initial gas pressure was estimated as $1.94\pm0.05\, mbar$ with the stopping model from Ref. \cite{bethe1930theorie,bloch1933bremsung}. In fact, the initial gas pressure was also checked by the measured energy loss of $100\,keV$ proton beam, the accuracy is better than $5\%$.

The temporal profile of the discharging current, measured by a Rogowsky coil, are also shown in Fig.2. They develop similar to each other since both the discharging current and the energy loss are mainly dependent on the free electron density in the plasma. As observed for the first $1$ microsecond, the discharge-current and the energy-loss are not very stable. It is probably due to the fast changing of the electromagnetic field in the beginning. When the discharging current reaches the maximum at around 3us after discharging, the temporal gradients of the electromagnetic field and the plasma parameters are minimum, and the free electron density in plasma reaches the maximum. The plasma parameters at the peak-discharging time are very well diagnosed, the linear free electron density and the ionization degree are $4.48\cdot 10^{17} /cm^{2}$ and $31\%$, respectively \cite{kuznetsov2013measurements}. So in the following we discuss the data at this time and compare it with our theoretical approaches and others.
\vspace{-0.5cm}
\begin{figure}[htb]
\setlength{\abovecaptionskip}{-0.0cm}
\centerline{\includegraphics[width=9.5cm]{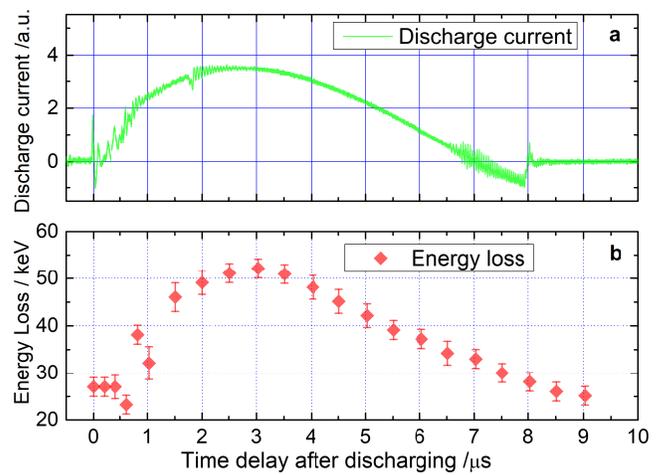}}
\caption{\label{fig:wide}Temporal evolution of discharging-current (${\mathbf{a}}$) and Energy-loss(${\mathbf{b}}$) of $100\,keV He$ ion in the plasma}
\setlength{\belowcaptionskip}{-0.2cm}   
\end{figure}

As is shown in Fig. 3, the measured energy loss (EXP., red dot) is more than $30\%$ higher than the predictions of the semi-classical approaches [28,29] if taking the effective charge of $1.43$ from the empirical formula of Ref. \cite{gus2009method} ($T\_Z_{eff}$,green triangle). By solving the Time Dependent Rate Equation (TDRE) with all the main projectile excited states and for all relevant atomic processes, our calculations (TDRE\_A,blue square) are in excellent agreement with the experimental data. In order to show the importance of the excited states and radiation decay in the stopping process, we also made the calculations for the cases that only ground states (TDRE\_G, cray diamond) are included or only Radiation-decays are excluded (TDRE\_R, purple star) in our model.

\begin{figure}[htb]
\setlength{\abovecaptionskip}{-0.3cm}
\centerline{\includegraphics[width=9.5cm]{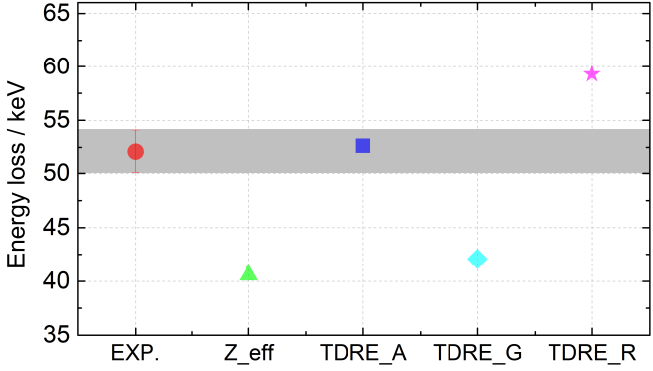}}
\caption{\label{fig:wide}Comparison between the measured energy loss (EXP.) and the theoretical predictions from different models. Here Z\_eff stands for the simi-classical calculation with effective charge from the empirical formula in Ref. \cite{gus2009method}, others represent the predictions in case all the main atomic states and atomic processes are included (TDRE\_A), or only ground states (TDRE\_G) are included or only Radiation-decays are excluded (TDRE\_R) in our model }
\setlength{\belowcaptionskip}{-0.2cm}   
\end{figure}

Since the nuclear stopping is neglect, the stopping process of the $100\,keV/u$ helium ion in the hydrogen plasma mainly contents two parts: stopping caused by collisions with bound electrons ($S_{b}$) and that with free electrons ($S_{f}$). In our model, $S_{b}$ is calculated by classical trajectory Monte Carlo method \cite{olson1977charge,he2009charge} in an ab initio and self-consistent way. These calculations are very reliable and agree well with the experimental data \cite{allison1958experimental} and theoretial predictions \cite{ziegler1999stopping}. $S_{f}$ is calculated by the dielectric-response theory \cite{brandt1982effective}:
\begin{equation}
{S_f} = \frac{{{e^2}}}{{\pi v_p^2}}\int_{{\kern 1pt} 0}^{{\kern 1pt} \infty } {\frac{{dk}}{k}{{\left| {{Z_p} - \rho (k)} \right|}^2}} \int_{{\kern 1pt}  - k{v_p}}^{{\kern 1pt} {\kern 1pt} k{v_p}} {d\omega \;\omega } \;{\rm{Im}}\left[ { - \frac{1}{{\varepsilon (k,\omega )}}} \right]
\end{equation}
where $\varepsilon(k,\omega)$ is the dielectric function, $Z_{p}$ is the projectile nuclear charge, and $¦Ñ(k)$ is the Fourier transform of the projectile bound electron density, which is depending on the electronic configuration (or atomic state) of the projectile. The stopping power is then obtained by summing over both $S_{b}$ and $S_{e}$ for all the projectile atomic states, as following
\begin{equation}
S = \sum\limits_i {P(i)S(i)}  = \sum\limits_i {\left( {P(i){S_b}(i){n_b} + P(i){S_f}(i){n_f}} \right)}
\end{equation}
where $P(i)$ denotes the fraction of the projectiles in atomic state $i$, nb and ne is the density of bound electron and free electron respectively. The energy loss of the ion in plasma is calculated by integrating the above stopping along the projectile's trajectory.

Instead of taking an effective-charge ($Z_{eff}$) in the semi-classic theory \cite{sigmund2001effective}, this model takes account the contribution of all the projectile atomic state which is more objective to approach the real interaction of ion and plasma. In our case, a total of ten significant configurations, i.e. $He^{2+}, 1s,2s,2p,3s,3p,3d $ for $He^{+}$, and $1s^{2},1s2s,1s2p$ for $He$. In the simulation those configurations with high excite states are considered to be $3d$ for $He^{+}$ and $1s2p$ for $He$, respectively.

In order to get the fraction of the projectile atomic state, the following time dependent rate equation (TDRE) are solved,
\begin{equation}
\frac{{d{P_i}({v_p},t)}}{{dt}} = \sum\limits_{j( \ne i)} {[\alpha (j \to i){P_j}({v_p},t) - \alpha (i \to j){P_i}({v_p},t)]}
\end{equation}
where $P_{i}(v_{p},t)$ denotes the time ($t$) dependent fraction of the projectiles in configuration $i$, which is related on the projectile velocity $v_{p}$, $\alpha(j\rightarrow i)$ is the rate coefficient for the transferring from the atomic state $j$ to $i$. All the main atomic processes like collisional ionization, collisional excitation and the respective reverse processes like radiative decay (RD), radiative recombination (RR), dielectric recombination (DR) and 3-body recombination (3BR) are included in the equation. The cross-sections for the atomic collisions between the projectile and the free electrons, bound electrons and protons in the plasma are either calculated by solving the time dependent Schr{\H o}dinger equation \cite{chun2015charge}or using the Flexible Atomic Code from Ref. \cite{gu2008flexible}, or from the data in Ref. \cite{winter2007electron,haar1993measurement}. It needs to be noted that, under the experimental condition, the charge state and atomic states get quasi equilibrium in nanoseconds with the corresponding equilibrium length of about $5-10\,mm$ which is much shorter than the interaction length.
\begin{figure}[htb]
\setlength{\abovecaptionskip}{-0.1cm}
\centerline{\includegraphics[width=9.5cm]{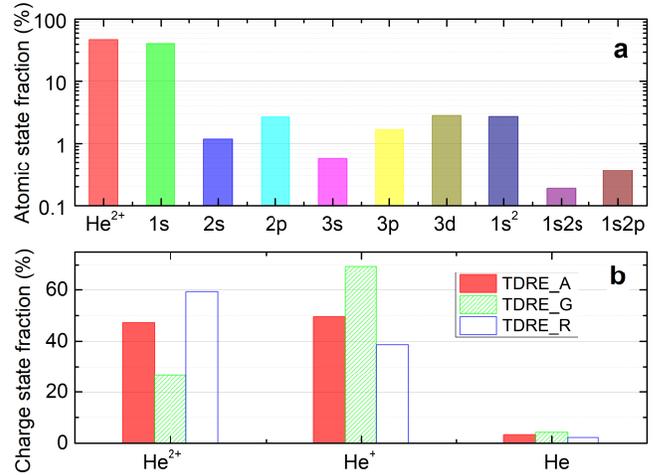}}
\caption{\label{fig:wide}Atomic state ($\mathbf{a}$) and charge state (${\mathbf{b}}$) distributions for the hellion ion in plasma. TDRE-A and TDRE-G are the same as those in Fig. 3.}
\setlength{\belowcaptionskip}{-0.2cm}   
\end{figure}

Figure 4 shows the atomic state distribution ($\mathbf{a}$) and charge state distribution ($\mathbf{a}$) of $100\,keV/u$ hellion ion in the plasma, which is calculated by TDRE as Eq. (3). The atomic state distribution is calculated by TDRE which includes all the 10 atomic states and the related main atomic progress (TDRE\_A), the corresponding charge state distribution for $He^{+}$ and $He$ is the sum over all their corresponding states. It is found in Fig.4($\mathbf{a}$) that, $He^{2+}$ and the ground state of $He^{+}$ i.e. $1s$ state, are the dominate atomic states, while the total fraction of all the 8 excited states is less than $10\%$. However, if we do not include all the main excited state in the calculations as shown in Fig.4 ($\mathbf{b}$) TDRE\_G, the charge state distribution will be changed significantly. It is interesting that, our TDRE\_A calculation suggests a mean charge state of 1.44, which is very close to the effective charge state (1.43) deduced from the empirical formula in Ref. \cite{gus2009method}, but the effective charge description lead to a wrong result which predict a much lower energy-loss (Fig.3). If all the projectile electrons are assumed to be in the ground states, the TDRE\_G model gives an estimation of a much smaller effective charge of 1.17, which fails to explain the energy loss as well. If radiative decay is artificially suppressed (TDRE\_R) in the model, it will lead to a remarkable increasing of the mean charge state and an unacceptable over-estimation of the energy loss. Therefore it is concluded that the projectile excited states play an important role in ion stopping, even though all together 8 excited states have only a fraction of less than $10\%$.

Since a number of atomic processes occur, including collisions with the free electron, bound electron and the hydrogen ions, it is necessary to study the role of each process with our model. Based on our calculation, it is found that the excited states may act as a ¡°setter¡± to adjust the atomic state distribution, since the rate coefficient for the excited states could be much higher than that for the ground states. For instance, the cross-section for a $He^{2+}$ ion to capture an electron to its excited states is much higher than that to its ground state, the radiative decay of the excited $He^{+}$ contribute the most part of the ground state of $He^{+}$ due to their high rate coefficient, the impact ionization of the excited states of $He^{+}$ is comparable to that of the ground state due to their lower ionization energy, and the collisional excitation processes are also of the key transfer channels where existence of excited states are necessary as well.

In summary, the stopping of low energy alpha particle in a hydrogen plasma was investigated both experimentally and theoretically. By combining of the novel accelerator in $100\,keV/u$ energy regime with a well-characterised homogenously plasma, experimental data with very high precision has been achieved. The measured energy loss exceeds the prediction of the commonly used semi-classical models by about $30\%$, which means the fail of the effective-charge descriptions. Taking all the main projectile excited states and all the main atomic processes into account, our model based on the Time Dependent Rate Equation fit excellent with the measurements. It is revealed that the active involvement of excited states in the charge transfer processes may have a remarkable influence to the atomic state evolution and the stopping of the low energy ions in plasma. The experimental data and the certified theoretical methods provide an important support for the relevant research like DT-alpha heating, atomic process in solar-wind and for the application of plasma strippers in modern accelerators.

\section*{Acknowledgments}

We sincerely thank the Jinyu Li and his co-workers for running the HIRFL accelerator. This work was supported by the National Key R $\&$ D Program of China (Grant Nos. 2017YFA0402300 and 2019YFA0404900), the Chinese Science Challenge Project No. TZ2016005, and the National Natural Science Foundation of China (Grants Nos. U1532263, 11574034, 11705141, 11775282, and U1530142).


\bibliography{ref}

\end{document}